\documentclass[aps,prl,twocolumn,superscriptaddress,showpacs]{revtex4}
\usepackage{graphicx}% Include figure files
\usepackage{dcolumn}% Align table columns on decimal point
\usepackage{bm}% bold math
\usepackage{verbatim}
\usepackage{color}
\usepackage[colorlinks]{hyperref}
\input{epsf}
\begin{document}

%\preprint{APS/123-QED}

\title{Transport study of Berry's phase, the resistivity rule, and quantum Hall effect in graphite}

\author{Aruna N. Ramanayaka}
% \altaffiliation{Department of Physics and Astronomy, Georgia State University,
%Atlanta, GA 30303.}%Lines break automatically or can be forced with \\
\author{R. G. Mani}
\affiliation{Department of Physics and Astronomy, Georgia State
University, Atlanta, GA 30303.}

\date{\today}% It is always \today, today,

\begin{abstract}
Transport measurements indicate strong oscillations in the Hall-,
$R_{xy}$, and the diagonal-, $R_{xx}$, resistances and exhibit
Hall plateaus at the lowest temperatures, in three-dimensional
Highly Oriented Pyrolytic Graphite (HOPG). At the same time, a
comparative  Shubnikov-de Haas-oscillations-based Berry's phase
analysis indicates that graphite is unlike the GaAs/AlGaAs 2D
electron system, the 3D n-GaAs epilayer, semiconducting
$Hg_{0.8}Cd_{0.2}Te$, and some other systems. Finally, we observe
the transport data to follow $B\times dR_{xy}/dB \approx - \Delta
R_{xx}$. This feature is consistent with the observed relative
phases of the oscillatory $R_{xx}$ and $R_{xy}$.
\end{abstract}

\maketitle

%\section{introduction}
Single layers of carbon atoms known as graphene are an interesting
future electronic material.\cite{grid1,grid2,grid3,grid4} At the
same time, graphene is a novel Two-dimensional Electronic System
(2DES) with remarkable features providing for a solid state
realization of quantum electrodynamics, massless Dirac particles,
an anomalous Berry's phase,\cite{grid5, grid6, grid7} and an
unconventional quantum Hall effect in a strong magnetic
field.\cite{grid1,grid2,grid3,grid4,grid5,grid6,grid7} The $ABAB$
bernal stacking of graphene helps to produce graphite, an
anisotropic electronic material exhibiting a large difference
between the in-plane and perpendicular transport. Since graphite
may be viewed as stacked graphene, one wonders whether remnants of
the remarkable properties of graphene might also be observable in
graphite.

Graphite exhibits a $\approx0.03 eV$ band overlap, in contrast to
the zero-gap in graphene.\cite{grid8,grid9,grid10} In the recent
past, concepts proposed for graphene have also been invoked for
graphite. For example, ref.\cite{grid11} argues for the
observation of massive majority electrons with a three dimensional
(3D) spectrum, minority holes with a two dimensional (2D)
parabolic massive spectrum, and majority holes with a 2D Dirac
spectrum. Hall plateaus in $\sigma_{xy}$ extracted from
van-der-Pauw measurements on highly oriented pyrolytic graphite
(HOPG) have been cited as evidence for the integral quantum Hall
effect (IQHE) in graphite in ref. \cite{grid12}. Luk'yanchuk et
al. in ref. \cite{grid13} suggest simultaneous quantum Hall
effects for the massive electrons and massless Dirac holes with
Berry's phase of $\beta = 0$ and $\beta = 1/2$, respectively.
$\beta$ is given here in units of $2\pi$. Finally, an ARPES-study
has reported massless Dirac fermions coexisting with finite mass
quasiparticles in graphite.\cite{grid14}

Plateaus at Hall resistance at $R_{xy} = h/ie^{2}$ with $i =
1,2,3,...$ and $R_{xx} \rightarrow 0$ are typically viewed as a
characteristic of IQHE in the 2DES. At the same time, a "thick"
specimen consisting of quantum wells separated by wide barriers
exhibits plateaus at $R_{xy} = h/ije^{2}$, where $j$ counts the
number of layers in the specimen. Indeed, introducing a dispersion
in the z-direction does not modify the observability of
IQHE.\cite{grid15} Thus, IQHE can also be a characteristic of
anisotropic 3D systems such as graphite. Here, in graphite, the
IQHE can be of the canonical variety, or of the unconventional
type reported in graphene. In addition, theory has predicted the
existence of a single, true bulk 3D QHE in graphite in a large
magnetic field parallel to the c-axis at $\sigma_{xy} =
(4e^{2}/\hbar)(1/c_{0})$, where $c_{0} = 6.7\AA$ is the c-axis
lattice constant.\cite{grid16}

%%%%%%%%%%%%%%%%%%%%%%%%%%%%%%%%%%%%%%%%%%%%%%%%%%%%%%%%%%%%%%%%%%%%%%
\begin{table}
\caption{\label{tab:table1}Intercept $n_{0}$ and the slope $B_{0}$
of ($n$) vs. $B^{-1}$ plot. $\beta$ is the suggested Berry's phase
in units of $2\pi$.}
\begin{ruledtabular}
\begin{tabular}{lccc}

 Material &$n_{0}$ &$B_0 (Tesla)$ &$\beta$\\
\hline Graphite &$0.47 \pm 0.02$ & 4.52 & $1/2$\\ GaAs/AlGaAs
&$0.05 \pm 0.01$ & 6.45& $0$\\ GaAs\footnotemark[1] \cite{grid20}
&$0.06 \pm 0.02 $ & 7.43& $0$\\
$Hg_{0.8}Cd_{0.2}Te$\footnotemark[2] \cite{grid21} &$-0.003 \pm
0.022 $ & 1.298 & $0$\\ HgTe\footnotemark[3] \cite{grid22} & $0.06
\pm 0.03$ & 26.14& $0$\\ 3D AlGaN \cite{grid23} & $-0.01 \pm 0.03$
& 35.39& $0$\\ InSb \cite{grid24} & $0.05 \pm 0.03$ & 19.80& $0$\\
$C_{9.3}AlCl_{3.4}$\footnotemark[4] \cite{grid25} & $0.48 \pm
0.02$& 11.7 & $1/2$\\
\end{tabular}
\end{ruledtabular}
\footnotetext[1]{$2 \mu m$ thick n-GaAs epilayer}
\footnotetext[2]{bulk semiconducting $Hg_{0.8}Cd_{0.2}Te$}
\footnotetext[3]{n-type HgTe quantum wells} \footnotetext[4]{1st
stage graphite intercalation compound}
\end{table}
%%%%%%%%%%%%%%%%%%%%%%%%%%%%%%%%%%%%%%%%%%%%%%%%%%%%%%%%%%%%%%%%%%%%%%
Thus, there are reasons for carrying out quantum Hall transport
studies in graphite. The possibility of both the canonical and
unconventional IQHE in graphite motivates also a study of the
Berry's phase. Finally, one wonders whether 3D graphite might also
satisfy the resistivity rule observed in 2D quantum Hall
systems.\cite{grid17}

%\begin{comment}
\begin{figure}[t]
%h=here, t=top, b=bottom, p=separate figure page
\begin{center}
\leavevmode \epsfxsize=2.25 in \epsfbox {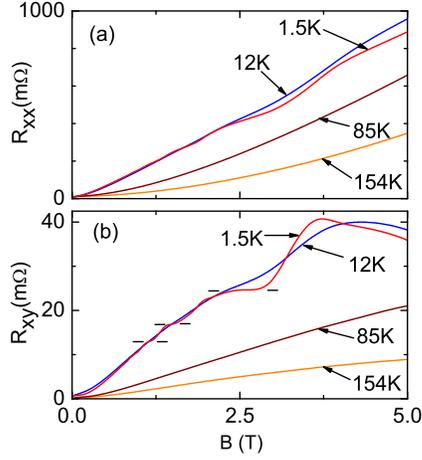}
\end{center}
%\begin{figure}
%\includegraphics{fig1}% Here is how to import EPS art
\caption{\label{fig:epsart} (color online). (a) The diagonal
resistance ($R_{xx}$) and (b) Hall resistance ($R_{xy}$) of sample
S1 are exhibited vs. the applied magnetic field, $B$, between $154
\leq T \leq 1.5 K$. $R_{xy}$ plateaus and SdH oscillations in
$R_{xx}$ are manifested at low $T$.}
\end{figure}

\begin{figure}[t]
%h=here, t=top, b=bottom, p=separate figure page
\begin{center}
\leavevmode \epsfxsize=2.25 in \epsfbox {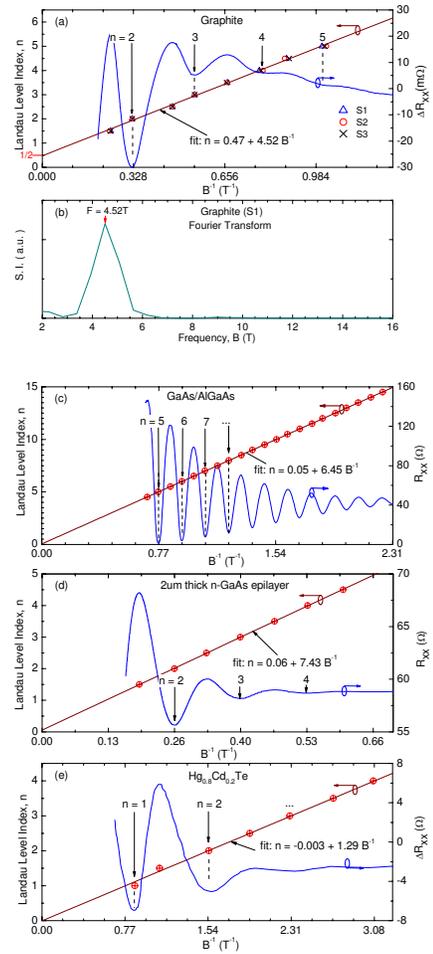}
\end{center}
%\begin{figure}
%\includegraphics{fig2}% Here is how to import EPS art
\caption{\label{fig:epsart} (color online).(a) Shown here are
plots of Landau level index ($n$)(left) of samples S1, S2, and S3,
and $R_{xx}$  (right) of sample S1, vs. $B^{-1}$. The slope of the
$n vs. B^{-1}$ plot indicates the SdH frequency, $F = 4.52 Tesla$,
and the intercept, $n (B^{-1} = 0) = 0.47$. (b) This plot shows
the spectral intensity of the Fourier transform of $\Delta
R_{xx}(1/B)$ of sample S1. A single peak in the Fourier spectrum
confirms that the SdH effect in these graphite specimens is
basically dominated by one type of carrier with $F = 4.52 Tesla$.
(c) $n$ and $R_{xx}$ are shown vs. $B^{-1}$ for the GaAs/AlGaAs 2D
electron system. In (d) \& (e) $n$ and $R_{xx}$ are shown for the
$2 \mu m$ thick n-GaAs epilayer and 3D $Hg_{0.8}Cd_{0.2}Te$
systems. Linear fit of $n$ vs. $B^{-1}$ intersects the ordinate at
$0.47$, $0.05$, $0.06$, and $-0.04$ for HOPG, GaAs/AlGaAs, $2 \mu
m$ thick n-GaAs epilayer, and bulk $Hg_{0.8}Cd_{0.2}Te$ systems,
respectively. }
\end{figure}

\begin{figure}[t]
%h=here, t=top, b=bottom, p=separate figure page
\begin{center}
\leavevmode \epsfxsize=2.25 in \epsfbox {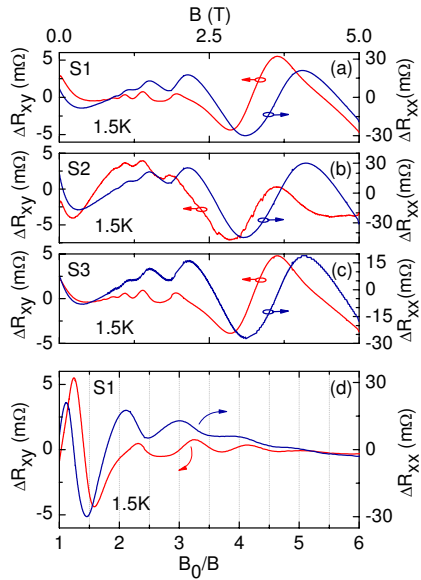}
\end{center}
%\begin{figure}
%\includegraphics{fig3}% Here is how to import EPS art
\caption{\label{fig:epsart} (color online). The oscillatory Hall-
($\Delta R_{xy}$) and diagonal- ($\Delta R_{xx}$) resistances are
shown for samples (a) S1, (b) S2, and (c) S3, vs. B. (d) $\Delta
R_{xy}$  and $\Delta R_{xx}$ are shown vs. the normalized inverse
magnetic field $B_{0}/B$. This plot shows a $\pi/2$ phase shift
between $\Delta R_{xx}$ and $\Delta R_{xy}$.}
\end{figure}

\begin{figure}[t]
%h=here, t=top, b=bottom, p=separate figure page
\begin{center}
\leavevmode \epsfxsize=2.25 in \epsfbox {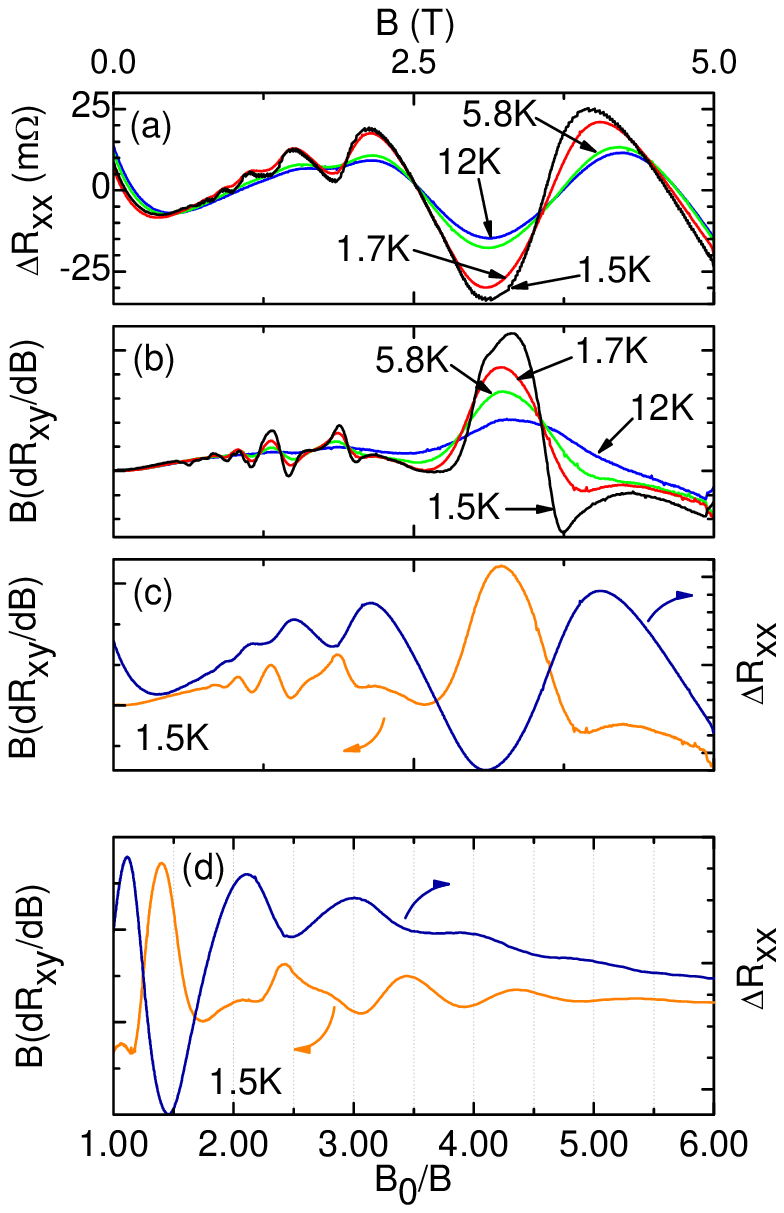}
\end{center}
%\begin{figure}
%\includegraphics{fig4}% Here is how to import EPS art
\caption{\label{fig:epsart} (color online). Panels (a) and (b)
show that the phase of the oscillatory $\Delta R_{xx}$ and
$B\times dR_{xy}/dB$ does not change with $T$. (c) Comparison of
$\Delta R_{xx}$ and $B\times dR_{xy}/dB$ at $T = 1.5 K$. (d)
$\Delta R_{xx}$ and $B\times dR_{xy}/dB$ are shown vs. $B_{0}/B$.
Both (c) and (d) suggest a phase shift of approximately $\pi$
between $\Delta R_{xx}$ and $B\times dR_{xy}/dB$.}
\end{figure}
%\end{comment}

This magneto-transport study of HOPG shows strong oscillations in
the Hall resistance, $R_{xy}$, and Shubnikov-de Haas (SdH)
oscillations in the diagonal resistance, $R_{xx}$, while
manifesting Hall plateaus at the lowest temperatures. A Fourier
transform of the $R_{xx}$ SdH oscillations indicates a single set
of carriers, namely electrons, in these HOPG specimens. A Berry's
phase analysis of the SdH data suggests that these carriers in
graphite are unlike those in GaAs/AlGaAs heterostructures, n-GaAs
epilayers, bulk semiconducting $Hg_{0.8}Cd_{0.2}Te$, the HgTe
quantum well, 3D AlGaN, and InSb systems. Further, a resistivity
rule study of graphite indicates $R_{xx} \sim - B \times
dR_{xy}/dB$, in variance with observations in canonical quantum
Hall systems, while a phase analysis suggests a new classification
(``type-IV")\cite{grid18} of the phase relation between the
oscillatory $R_{xx}$ and $R_{xy}$. The latter observation is
consistent with $R_{xx} \sim - B \times dR_{xy}/dB$.

Our $25 \mu m$ thick graphite specimens were exfoliated from bulk
HOPG, and the measurements were carried out using standard lock-in
techniques with the $B$ parallel to the c-axis. Measurements of
$R_{xx}$ and $R_{xy}$ are shown in Fig.~1(a) and (b) respectively
for $1.5 \leq T \leq 154 K$ and $ 0 \leq B \leq 5$ Tesla. At low
$T$, SdH oscillations in $R_{xx}$ and $R_{xy}$, and plateaus in
$R_{xy}$, appear manifested, see Fig.1(a) \& (b), as in Fig. 2 and
Fig. 3 of ref.\cite{grid19}. Three Hall plateaus are observable in
Fig. 1(b). The Hall plateau resistance observed at $B = 2.5T$ and
$T = 1.5K$ is consistent, within a factor-of-two, with viewing the
specimen as a stack of uncoupled quantum Hall layers. Note that
$R_{xx} > 0$ through the Hall plateaus. The finite tilt of the
highest-$B$ plateau at $T = 1.5K$ might be due to the incipient
$R_{xy}$ saturation, possibly due to multi-band transport.

The SdH effect has been used to probe the Berry's
phase.\cite{grid5, grid6} For the graphene system, the oscillatory
$R_{xx}$ has been written as $R(B,T)\cos(2\pi(B_0/B + 1/2 +
\beta))$\cite{grid7}, where $R(B,T)$ is the SdH oscillation
amplitude, $B_0$ is the SdH oscillation frequency, and $\beta$ is
the Berry's phase in units of $2\pi$. The observation of an
anomalous Berry's phase in graphene has generated new interest in
the study of the same in graphite. Hence, we set out to examine
our data for graphite from a similar perspective.

Graphite is a semi-metal with majority electrons and holes
exhibiting elongated Fermi surfaces along the c-axis.\cite{grid9}
The Fermi surface of graphite has been studied via the oscillatory
magnetization (de Haas -van Alphen [dHvA]) effect, and also the
oscillatory magnetoresistance (SdH) effect.\cite{grid9} Graphite
exhibits a strong dHvA effect, and the Fourier transform of the
associated oscillatory magnetization readily exhibits two spectral
peaks corresponding to the two sets of majority carriers. The SdH
data, on the other hand, can often exhibit just a single broad
peak in the Fourier spectrum.\cite{grid13} It has been suggested
that the scattering lifetimes manifested by the carriers in the
two experiments might not be the same.\cite{grid9} Further, that
one set of carriers might suffer more broadening, and as a
consequence, the associated oscillatory contribution due to this
band might vanish in the SdH effect.

The HOPG specimens examined in this study are also characterized
by a single broad peak in the Fourier transform of the SdH data,
see Fig. 2(b). This feature suggests that the SdH effect in our
HOPG specimen is also dominated by one type of carriers, namely
electrons, and this feature indicates that this single
carrier-type may be subjected to a Berry's phase analysis as in
graphene.

In order to examine and compare the Berry's phase, plots of the
Landau level index, $n$ (left axis), and $R_{xx}$ (right axis)
versus $B^{-1}$ are shown in Fig.~2(a), (c), (d), and (e) for
graphite, the GaAs/AlGaAs 2D electron system, an n-GaAs
epilayer\cite{grid20}, and bulk semiconducting
$Hg_{0.8}Cd_{0.2}Te$\cite{grid21} respectively. For all four
material systems, minima of oscillatory $R_{xx}$ have been
assigned with $n$, and maxima of the oscillatory $R_{xx}$ have
been assigned with  $n+1/2$, as in ref. \cite{grid7}. In such an
analysis, an intercept $n_{0} = 0$ would normally indicate a
Berry's phase $\beta = 0$, while $n_{0} = 1/2$ would normally
correspond to $\beta = 1/2$, as in graphene.\cite{grid7}. For the
three graphite samples S1, S2, and S3 (see Fig.~2(a)), the Landau
level index intercept of a linear fit to $n$ vs. $B^{-1}$ yields
$0.47$, i.e., $1/2$. Panels (c), (d), and (e) give intercepts of
$0.05$, $0.06$, and $-0.04$, i.e., zero, for the GaAs/AlGaAs,
n-GaAs epilayer, and bulk $Hg_{0.8}Cd_{0.2}Te$ systems. These
fit-extracted intercepts, $n_{0}$, and $B_{0}$ have been
summarized in Table 1.

A similar analysis was carried out with other data and these
results are also summarized in Table~\ref{tab:table1} for the HgTe
quantum well,\cite{grid22} 3D AlGaN,\cite{grid23}
InSb,\cite{grid14} and $C_{9.3}AlCl_{3.4}$ - a first stage
graphite intercalation compound.\cite{grid25} Notice that the
Table~\ref{tab:table1} represents eight different systems, yet
only graphite and $C_{9.3}AlCl_{3.4}$  show $n_{0} = 1/2$ in the
$B^{-1}\rightarrow 0$ limit as observed in graphene.\cite{grid7}
This is a principal experimental finding of this work.

Does this experimental finding signify an anomalous Berry's phase
for the observed carriers in graphite? If the experimental results
shown here in Fig. 1 had indicated $R_{xy} > R_{xx}$ as in ref. 7,
then, certainly, the conclusion would immediately follow that
electrons in graphite exhibit an anomalous Berry's phase as in
graphene. The fact that Fig. 1 indicates $R_{xy}<<R_{xx}$
complicates the issue since many would reason that
$R_{xy}<<R_{xx}$ implies $\rho_{xy} <<\rho_{xx}$ given the
geometrical factors in these graphite specimens, and as a
consequence, $n_{0}= 1/2$ in graphite should be taken to indicate
normal carriers,\cite{grid13} not Dirac carriers. Some features
should, however, serve as caution against following this line of
reasoning. First, strong SdH oscillations, as in Fig. 1, are
typically observed only when $\rho_{xy} > \rho_{xx}$, i.e.,
$\omega \tau > 1$, for the associated carriers. Thus, perhaps, the
observation of $R_{xy} << R_{xx}$ in graphite need not imply
$\rho_{xy} << \rho_{xx}$. Next, it appears worth pointing out that
some experiments in the GaAs/AlGaAs system have clearly shown that
there need not be a simple relation between $R_{xx}$ and
$\rho_{xx}$.\cite{grid27} Finally, it is also known that certain
types of longitudinal specimen thickness variations, density
variations, and/or current switching between graphene layers in
graphite, can introduce a linear-in-B component into $R_{xx}$
originating from the Hall effect; such an effect can produce
experimental observations of $R_{xx} >> R_{xy}$ even in systems
that satisfy $\rho_{xx} < \rho_{xy}$. Due to such possibilities in
graphite, the results shown Fig.~2 and Table 1 could plausibly
identify an anomalous Berry's phase for electrons in graphite, and
$C_{9.3}AlCl_{3.4}$ \cite{grid25}. Here, it is noteworthy that
semiconducting $Hg_{0.8}Cd_{0.2}Te$ specimen also exhibits $R_{xx}
> R_{xy}$\cite{grid21} and yet shows $n_{0}=0$, unlike graphite.
Clearly, the experimental finding is unambiguous: the infinite
field phase extracted from the SdH oscillations for graphite is
unlike the results for a number of canonical semiconductor
systems, see Table 1.

Next, we examine more closely the phase relation in the
oscillations of $R_{xx}$ and $R_{xy}$ since this is a striking
feature of the data. Thus, background subtracted $R_{xx}$ and
$R_{xy}$, i.e., $\Delta R_{xx}$ and $\Delta R_{xy}$, are shown in
the Fig.~3(a)- Fig.~3(c), for samples S1 - S3. Thus far, the phase
relations between $\Delta R_{xx}$ and $\Delta R_{xy}$ have been
classified into three types in the high mobility GaAs/AlGaAs
system.\cite{grid18} Namely, type-III, where the oscillations of
$\Delta R_{xx}$ and $\Delta R_{xy}$ are approximately $180$
degrees out of phase. Type-II, where the oscillation of the
$\Delta R_{xx}$ and $\Delta R_{xy}$ are in-phase. And, type-I,
where the peaks of $\Delta R_{xx}$ occur on the low-$B$ side of
the $\Delta R_{xy}$ peaks, with a $\pi/2$ phase shift. Fig.~3
shows, however, that the peaks of $\Delta R_{xx}$ appear on the
high-$B$ side of the $\Delta R_{xy}$ peaks, unlike the cases
mentioned above. This suggests a fourth phase relation
(``type-IV") between $\Delta R_{xx}$ and $\Delta R_{xy}$ in (HOPG)
graphite. In the $B^{-1}$ plot, see Fig.~3(d), the peaks of
$\Delta R_{xx}$ are shifted towards the low $B_{0}/B$ side with
respect to the peaks of $\Delta R_{xy}$, with an approximately
$\pi/2$ phase shift between $\Delta R_{xx}$ and $\Delta R_{xy}$,
confirming this conjecture.

For the resistivity rule study,\cite{grid17, grid18, grid28,
grid29} the temperature dependence of $\Delta R_{xx}$ and $B\times
dR_{xy}/dB$ are shown in Fig.~4(a) and Fig.~4(b) respectively.
Fig.~4 indicates a progressive change, where the oscillatory
amplitude increases with decreasing $T$ while the $B$-values of
the extrema remains unchanged with $T$. Fig.~4(c) shows the
$\Delta R_{xx}$ (right axis) and $B\times dR_{xy}/dB$ (left axis)
for the sample S1 at $1.5 K$. A direct comparison  (see Fig.~4(c))
reveals a $\pi$ phase difference between $B\times dR_{xy}/dB$ and
$\Delta R_{xx}$, i.e, $B\times dR_{xy}/dB \approx - \Delta
R_{xx}$. Plots of $\Delta R_{xx}$ and $B\times dR_{xy}/dB$ vs.
$B_{0}/B$, (see Fig.~4(d)) confirm an approximately $\pi$ phase
shift between $\Delta R_{xx}$ and $B\times dR_{xy}/dB$. This
variance from the canonical resistivity rule is consistent with
the observed Type-IV phase relation between the oscillatory
$R_{xx}$ and $R_{xy}$, see Fig.~3.

In summary, a study of (HOPG) graphite indicates strong
oscillatory $R_{xy}$ and $R_{xx}$, with Hall ($R_{xy}$) plateaus
coincident with a non-vanishing $R_{xx}$ at the lowest $T$. On the
other hand, a comparative SdH Berry's-phase study of graphite with
other systems such the GaAs/AlGaAs 2D electron system, the 3D
n-GaAs epilayer, bulk $Hg_{0.8}Cd_{0.2}Te$, HgTe quantum well, 3D
AlGaN, InSb, and $C_{9.3}AlCl_{3.4}$ - a graphite intercalation
compound- reveals a value of $n_{0}= 1/2$ in the
$B^{-1}\longrightarrow 0$ limit only for the graphene based
systems. As in graphene, this feature might indicate that one-half
of a conduction band Landau level is pinned together with one-half
of a hole Landau level in graphite and might reflect a non-trivial
Berry's phase ($\beta = 1/2$) and Dirac carriers in graphite. A
close study of the oscillatory diagonal- and Hall- resistances
reveals also an anomalous (``type-IV") phase relation between
oscillatory $R_{xx}$ and $R_{xy}$, over the entire $T$-range. This
result is consistent with the observation of $B\times dR_{xy}/dB
\approx - \Delta R_{xx}$ in the resistivity rule analysis of the
transport.

Work has been supported by the ARO under W911NF-07-01-0158, and by
the DOE under DE-SC-0001762. We acknowledge discussions with B.
Kaviraj and T. Ghanem.

\pagebreak
\begin{comment}
%%%%%%%%%%%%%%%%%%%%%%%%%%%%%%%%%%%%%%%%
\begin{figure}[p]
%h=here, t=top, b=bottom, p=separate figure page
\begin{center}
\leavevmode \epsfxsize=3.25 in \epsfbox {fig1eps.eps}
\end{center}
%\begin{figure}
%\includegraphics{fig1}% Here is how to import EPS art
%\caption{\label{fig:epsart}}
\end{figure}

%\pagebreak
\begin{figure}[p]
%h=here, t=top, b=bottom, p=separate figure page
\begin{center}
\leavevmode \epsfxsize=3.25 in \epsfbox {fig2eps.eps}
\end{center}
%\begin{figure}
%\includegraphics{fig2}% Here is how to import EPS art
%\caption{\label{fig:epsart}}
\end{figure}

%\pagebreak
\begin{figure}[p]
%h=here, t=top, b=bottom, p=separate figure page
\begin{center}
\leavevmode \epsfxsize=3.25 in \epsfbox {fig3eps.eps}
\end{center}
%\begin{figure}
%\includegraphics{fig2}% Here is how to import EPS art
%\caption{\label{fig:epsart}}
\end{figure}

%\pagebreak
\begin{figure}[p]
%h=here, t=top, b=bottom, p=separate figure page
\begin{center}
\leavevmode \epsfxsize=3.25 in \epsfbox {fig4eps.eps}
\end{center}
%\begin{figure}
%\includegraphics{fig3}% Here is how to import EPS art
%\caption{\label{fig:epsart} }
\end{figure}

\end{comment}
%%%%%%%%%%%%%%%%%%%%%%%%%%%%%%%%%%%%%%%%

\end{document}